\documentstyle[twocolumn,epsf,aps,tabularx]{revtex}
\draft
\begin{document}
\title{Influence of extended dynamics on phase transitions in a driven 
lattice gas}

\author{Attila Szolnoki and Gy\"orgy Szab\'o}

\address{
Research Institute for Technical Physics and Materials Science \\
P. O. Box 49, H-1525 Budapest, Hungary}

\address{\em \today}

\address {
\centering {
\medskip \em
\begin{minipage}{15.4cm}
{}\qquad Monte Carlo simulations and dynamical mean-field approximations
are performed to study the phase transition in a driven lattice gas with 
nearest-neighbor exclusion on a square lattice. A slight extension of
the microscopic dynamics with allowing the next-nearest-neighbor
hops results in dramatic changes.
Instead of the phase separation into high- and low-density regions in
the stationary state the system exhibits a continuous transition belonging
to the Ising universality class for any driving.
The relevant features of phase diagram are reproduced by an
improved mean-field analysis.
\end{minipage}
}}

\maketitle

\narrowtext

The concept of universality is well established in equilibrium critical
phenomena. According to this concept only just a few parameters, i.e. spatial 
dimension of the system and the dimensionality of the order parameter 
determine the critical exponents meanwhile other details like the 
microscopic dynamics are irrelevant. 
The nonequilibrium systems exhibit richer and more complex
feature. One of the important questions to address is whether this concept
also applies to nonequilibrium systems. There are examples when a 
slight extension of microscopic dynamics results in different morphology of
the stationary state \cite{rutenberg,szol99}. 
The importance of the dynamics also manifests in other problems, such as
chemically reactive mixtures \cite{glotzer} or driven diffusive systems (DDS)
\cite{alexander} where the microscopic and the supposed macroscopic model
yield different morphologies. The former nonequilibrium system also exemplifies
that the resulting stationary state may differ significantly
from that of the corresponding equilibrium model \cite{possib}.

Very recently Dickman has introduced a simple driven lattice gas model
whith hard-core interaction between the particles which excludes the 
simultaneous occupation of the nearest neighbor sites on a square
lattice \cite{ronnne}. In the absence of driving this model is equivalent
to a thoroughly investigated equilibrium model discussed in connection with the
theory of melting \cite{Domb,Temp,Burl,GF}. For this particular interaction
the only (control) parameter of the model is the particle concentration.
The driven version of this nearest neighbor exclusion (NNE) model can be
related to some traffic and granular flow models as detailed by
Dickman \cite{ronnne}.
Despite its simplicity the driven NNE model shows remarkable nonequilibrium
behavior such as the coexistence of a low- and high-density phases and the
current may decrease in response to increased drive. The order-disorder phase
transition also changes from second to first order in the presence of drive. 
In this Brief Report we slightly modify the microscopic dynamics to explore 
the robustness of this behavior. We demonstrate that the coexistence of two 
phases disappears and the phase transition remains continuous for all drives 
by allowing the next-nearest-neighbor hops too. A more accurate version of
the dynamical mean-field theory predicts a phase diagram in good qualitative
agreement with the results of Monte Carlo (MC) simulations.

\begin{figure}
\centerline{\epsfxsize=8cm
            \epsfbox{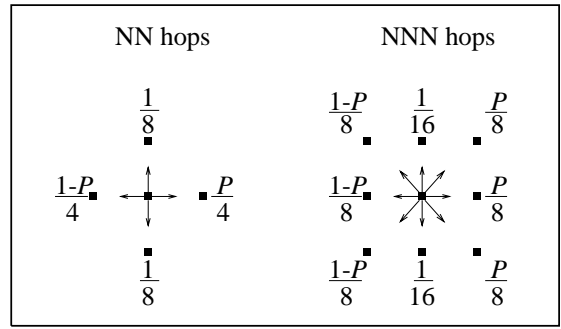}
            \vspace*{1mm} }
\caption{Schematic illustration of possible hopping rates of the 
nearest-neighbor (NN) and the extended next-nearest neighbor (NNN) 
hopping dynamics. The drive is horizontal and the directions are chosen
with equal probability [$1/4$ ($1/8$) for the NN (NNN) hops].}
\label{fig:hops}
\end{figure}

We consider two-dimensional driven lattice gases on a square lattice with 
$L \times L = N$ sites under periodic conditions. The occupation variables
$\sigma_i = 0 \,\, (1)$ if the lattice site $i$ is empty (occupied) and 
the concentration is defined as $\rho = \sum_i \sigma_i / N$. The only
interaction is to forbid the simultaneous nearest-neighbor occupancy.
During the time evolution a randomly chosen particle can hop to one 
of its empty neighboring sites satisfying the condition of NNE. 
In the basic model studied by Dickman the particles can hop only to the 
nearest neighbor sites while in this extended model the next-nearest-neighbor
hops are also permitted with a probability as defined in Fig.~\ref{fig:hops}. 
Here, the value of $P$ is
varied from $0.5$ to $1$. The case $P=0.5$ corresponds to the isotropic
hopping rate characteristic to the  equilibrium model and $P=1$
represents the infinitely strong drive. Further details and the review of 
the equilibrium properties can be found in Ref.~\cite{ronnne}.

In the stationary state of this model the particle distribution is
disordered if the concentration is low enough. For sufficiently high
concentration sublattice ordering can be observed, that is the particles
form a chequerboard-like pattern. In other words, if the square lattice
is divided into two interpenetrating sublattices ($A$ and $B$)
then the particles prefer to stay in one of these sublattices. The 
transition between the ordered and disordered states is characterized by
an order parameter 
\begin{equation}
\Phi = { \rho_A - \rho_B \over \rho_A + \rho_B } \,\,,
\label{eq:phi}
\end{equation}
where $\rho_A$ and $\rho_B$  denotes the concentration of particles in
sublattice $A$ and $B$. In the disordered phase $\Phi = 0$. Conversely,
$\Phi = 1$ (or $-1$) if all the particles are positioned in the sites of
sublattice $A$ ($B$). 
When decreasing the concentration the equilibrium system ($P=0.5$)
undergoes an order-disorder transition which is continuous and belongs
to the Ising universality class \cite{GF}.
In the driven system, however, this transition becomes first order
if only nearest neighbor hops are permitted \cite{ronnne}. The present
dynamics conserves the number of particles therefore the first order
transition is accompaned with the coexistence of high- and low-density 
phases.

To explore the phase diagram of the extended model, we have performed a
dynamical mean-field analysis. The most relevant details of this laborious 
technique are given in the Appendix of the work by Dickman \cite{ronnne}
at the levels of two- and four-site clusters. Using this method one can
determine the probability of possible configurations on a given set of 
clusters if the particle distribution is homogeneous. Similarly to the
NN hop model, the two- and four-site approximations are not capable to 
describe the effect of driving as a consequence of the strong constraints 
of NNE. This means that the results are independent of $P$ and equivalent 
to those obtained by using the tradional cluster variation method 
\cite{CVM,woodb} for the equilibrium system ($P=1/2$). 
The predictions for the phase transition point are $\rho_c^{2p}=0.25$ at 
two-point and $\rho_c^{4p}=0.317$ at four-point level. 

\begin{figure}
\centerline{\epsfxsize=8cm
            \epsfbox{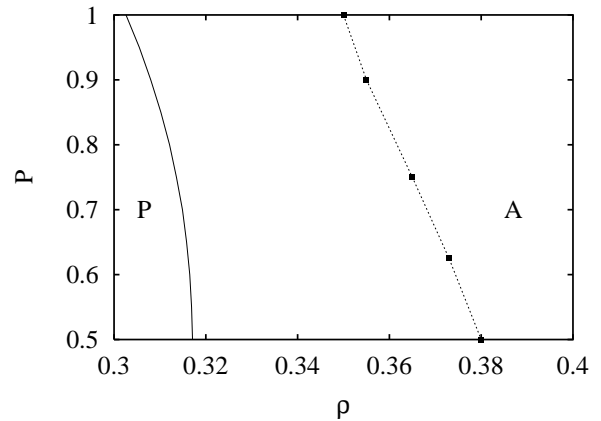}
            \vspace*{1mm} }
\caption{Phase diagram in the $\rho$ - $P$ plane. Solid curve is the
prediction of six-point mean-field approximation. Symbols represent Monte 
Carlo results. Dashed line is guide to eyes. The antiferromagnetic (A), and 
the paramagnetic (P) phases
are indicated.}
\label{fig:phd}
\end{figure}

The failure of dynamical mean-field approximation is disappointing since
previously this method gave qualitatively good phase diagrams for several
nonequlibrium models \cite{dmf}.
To overcome this shortage we have performed a six-point approximation 
used succesfully for a similar driven lattice gas \cite{6p}.
In this case we determine the configuration probabilities on $2 \times 3$
clusters. Taking the compatibility conditions and symmetries into account
the configuration probabilities are described by 17 parameters whose value
are determined by solving numerically the corresponding set of equations
of motion. At $P=1/2$ this calculation has reproduced the same solution
obtained at the level of four-point approximation. For the driven cases,
however, the solutions (for $\Phi=0$ and $\Phi > 0$) are already affected
by the value of $P$. As an example, for infinite drive the transition
appears at $\rho_c^{6p}(P=1)=0.3026$. The resulting phase diagram is shown
in Fig.~\ref{fig:phd}. We should mention that the phase transition remains
continuous for any drives according to this level of approximation.

To check these predictions MC simulations are performed for 
$L=20$, $40$, $80$, $160$, and $320$ under periodic boundary conditions
varying the concentration $\rho$ and drive $P$. 
To reach the stationary state we have used both homogeneous random and 
ordered initial states. In the former initial state the particles are
positioned within a strip parallel to the drive. 

In agreement with the expectation the modification of dynamics does not 
influence the 
stationary state if $P=0.5$ (equilibrium model). The MC data of $\Phi(\rho)$
functions collapse when comparing the results obtained for the NN and
NNN dynamics at any system sizes. At the same time the evolution
toward the stationary states becomes faster when NNN hops are allowed.
Similar effect was also reported in another driven diffusive model
\cite{rutenberg} as well as in domain growth processes \cite{ole}.
In contrast, a significant difference can be observed in
the stationary states when the systems are driven. While the separation of
the high- and low-density phases characterizes the NN dynamics \cite{ronnne}
leading to a jammed ``herring-bone'' structure
then the stationary state remains homogeneous for any drive and concentration
when NNN jumps are also allowed. The study of order parameter and density 
profiles
as a function of transversal coordinate also support that the ordered phase
is homogeneous in the case of extended dynamics. As a consequence,
the current has no size-dependence and varies smoothly with the 
concentration and $P$. 
The phase diagram obtained by MC simulations confirm the qualitative 
prediction of the dynamical mean-field theory. Namely, the critical
concentration ($\rho_c$) decreases with $P$ and the transition remains
continuous for those drives displayed in Fig.\ref{fig:phd}. 

\begin{figure}
\centerline{\epsfxsize=8cm
            \epsfbox{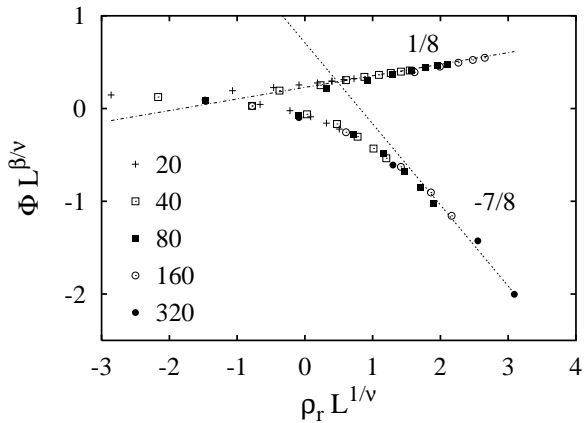}
            \vspace*{1mm} }
\caption{The finite-size scaling of order parameter of NNE model with NNN 
hopping for $P=1$. The slopes of the asymptote lines are indicated.}
\label{fig:scaling}
\end{figure}

The classification of the critical behavior for a nonequilibrium model
requires careful analysis. The difficulties are well demonstrated by
the investigation of standard model \cite{kls} whose critical behavior
is even controversial \cite{pedro}. Here, we present a finite-size scaling
analysis for the infinite strong drive ($P=1$) in the case of extended
(NNN) dynamics. For this purpose we have used the scaling form \cite{binder} 
\begin{equation}
\Phi(L,\rho) = L^{- \beta / \nu}\,\, \overline{\Phi} (\rho_r L^{1 / \nu})\,\,,
\label{eq:fss}
\end{equation}
where $\rho_r \equiv (\rho-\rho_c)/\rho_c$ is the reduced concentration.
Assuming that the nonequlibrium phase transition belongs to the class
of the equilibrium model (i.e. Ising exponents are supposed), nice data
collapse is found (see Figure~\ref{fig:scaling}). Here, 
$\rho_c^{MC}(P=1) = 0.350(5)$. This value is consistent with an alternating
estimate which comes from the analysis of the fourth order cumulant
of order parameter 
$U_L = 1 - {\langle \Phi^4 \rangle}_L /3 {\langle \Phi^2 \rangle}^2_L$ 
\cite{binder}. 
Similar critical behavior is found for any $P<1$ drive.

In summary, the microscopic dynamics has been slightly extended in a 
previously introduced driven diffusive lattice gas where the only interaction
is to forbid the simultaneous nearest-neighbor occupancy. 
Despite the weak modification the nonequlibrium behavior changes 
significantly. The stationary states are found to be homogeneous for any 
drives and densities if the next-nearest-neighbor hops are also permitted. 
In this case the system exhibits a sublattice ordering when the particle 
concentration is increased. The value of critical concentration decreases
with the drive and this behavior can be explained by the higher level of 
dynamical mean-field approximation. The numerical results support that the 
system is in the same universality class of the equilibrium Ising model.
This model exemplefies that the nonequlibrium behavior may be significantly
influenced by a weak modification of microscopic dynamics.

\vspace{0.5cm}

The authors thank Ron Dickman for stimulating discussions. This research was 
supported by the Hungarian National Research Fund (OTKA) under Grant No. 
F-30449  and T-33098. A. S. thanks the HAS for financial support by a Bolyai 
scholarship.

\end{document}